\documentclass[%
 reprint,
 amsmath,amssymb,
 aps,
]{revtex4-2}

\usepackage{color}
\usepackage{graphicx}
\usepackage{dcolumn}
\usepackage{bm}

\usepackage{graphicx}
\usepackage{dcolumn}
\usepackage{bm}
\usepackage{comment}
\usepackage{appendix}
\usepackage{physics}
\usepackage{soul}
\usepackage{url}

\begin{document}

\title{
Lattice-Trapped Atom Interferometry with a Bose-Einstein condensate:\\ Observation and Control of Interactions
}

\author{Emmett Hough} 
\email{emmetth@uw.edu}
\author{Tahiyat Rahman}
\author{Forest Tschirhart}
\author{Subhadeep Gupta}
\affiliation{Department of Physics, University of Washington, Seattle, WA, USA}

\date{\today}

\begin{abstract}
Precision interferometry with atomic wavepackets confined in a one-dimensional optical lattice is an emergent paradigm in quantum sensing of forces and fields, with applications in gravimetry, accelerometry, geophysics, and fundamental physics tests. We report on the realization of a lattice-trapped interferometer where the two arms are sourced from a weakly-interacting ytterbium Bose-Einstein condensate, coherently split and trapped by pulsed optical standing waves before recombination. We directly observe atomic interactions through contrast changes and phase shifts of the interferometer. By changing either the atom number or the sample volume to vary the density, we demonstrate control over interactions and optimize interferometer performance. Our observations are effectively captured by a mean-field theoretical model of the system. This work experimentally probes the boundary where improved performance from source brightening through higher phase space density transitions into a regime beyond single-atom physics in lattice-trapped atom interferometry, and opens a door to incorporating many-body effects for metrological advances in such platforms.
\end{abstract}

\maketitle

\section{Introduction}

Atom interferometers are well-established quantum sensors for fundamental physics such as tests of the standard model \cite{more20, abe21} and the equivalence principle \cite{asen20, dobk25, danf26}, as well as for applications such as gravimetry \cite{pand24, bena24}, gradiometry \cite{mcgu02, zhan23}, inertial sensing \cite{dick13,stol25}, navigation \cite{gers25} and geophysics \cite{bide18, xuej19}. The sensitivity of this technique improves with increased interferometer path separation and interrogation time while maintaining quantum coherence. The conventional paradigm of atom interferometer operation centers around freely-falling atoms where splitting, redirection, and recombination of the wavepackets are engineered with atom optics pulses, with the interferometer paths accumulating phase freely otherwise. Such free-space interferometers are ultimately limited in sensitivity by the finite spatial extent of the experimental apparatus, leading to second-scale interrogation times in the largest in-lab atomic fountain geometries \cite{asen17}. While major engineering efforts are being undertaken to extend these interrogation times with very large baseline atom interferometry \cite{abda25}, an appealing alternative is provided by trapped atom interferometers \cite{mcdo13, krzy23, chai26, lede25, ball24, char12, zhan16, xu19} where the force of gravity is canceled by an external potential, allowing extended interrogation times in table-top experiments. In particular, one-dimensional optical lattices with atomic wavepacket components trapped in spatially separated regions is emerging as a promising trapped atom interferometry platform \cite{char12, zhan16, xu19, pand24, ball24} with recent achievement of impressively long minute-scale coherence times \cite{pand24, pand24_2}. While interrogation times and path separations may be extendable with colder and denser samples, the consequent growth of interatomic interactions is crucial to address \cite{hori06,burc20}.

Here we introduce Bose-Einstein condensates (BECs) of ytterbium atoms as a bright coherent source for lattice-trapped atom interferometry. This limiting case of near-monochromatic velocity distribution allows us to observe atomic interaction effects directly in interferometer contrast variation and phase shift. We optimize the contrast through control of source expansion time prior to trapping the matterwave components during interrogation. Both the contrast variation and phase shift are effectively captured through a mean-field model of the experiment. We also find the interferometer performance for a BEC source to exceed a thermal source. These results benchmark the observation and control of interaction effects in the emerging paradigm of trapped atom interferometry and provide a launching point for metrology advancement using many-body effects. 

The remainder of this paper is organized as follows. In Secs.~\ref{sec:scheme} and ~\ref{sec:exp_setup} we introduce the interferometer scheme and provide details about the experimental setup. In Sec.~\ref{sec:MFD} we present the observation and control of interaction effects in our interferometer along with a mean-field theoretical model, followed by a discussion of the optimized performance in Sec.~\ref{sec:IfmPerf}. We present our conclusions and outlook in Sec. VI.

\section{Interferometer Scheme}\label{sec:scheme}

\begin{figure}[t]
     \centering
     \includegraphics[width=\linewidth]{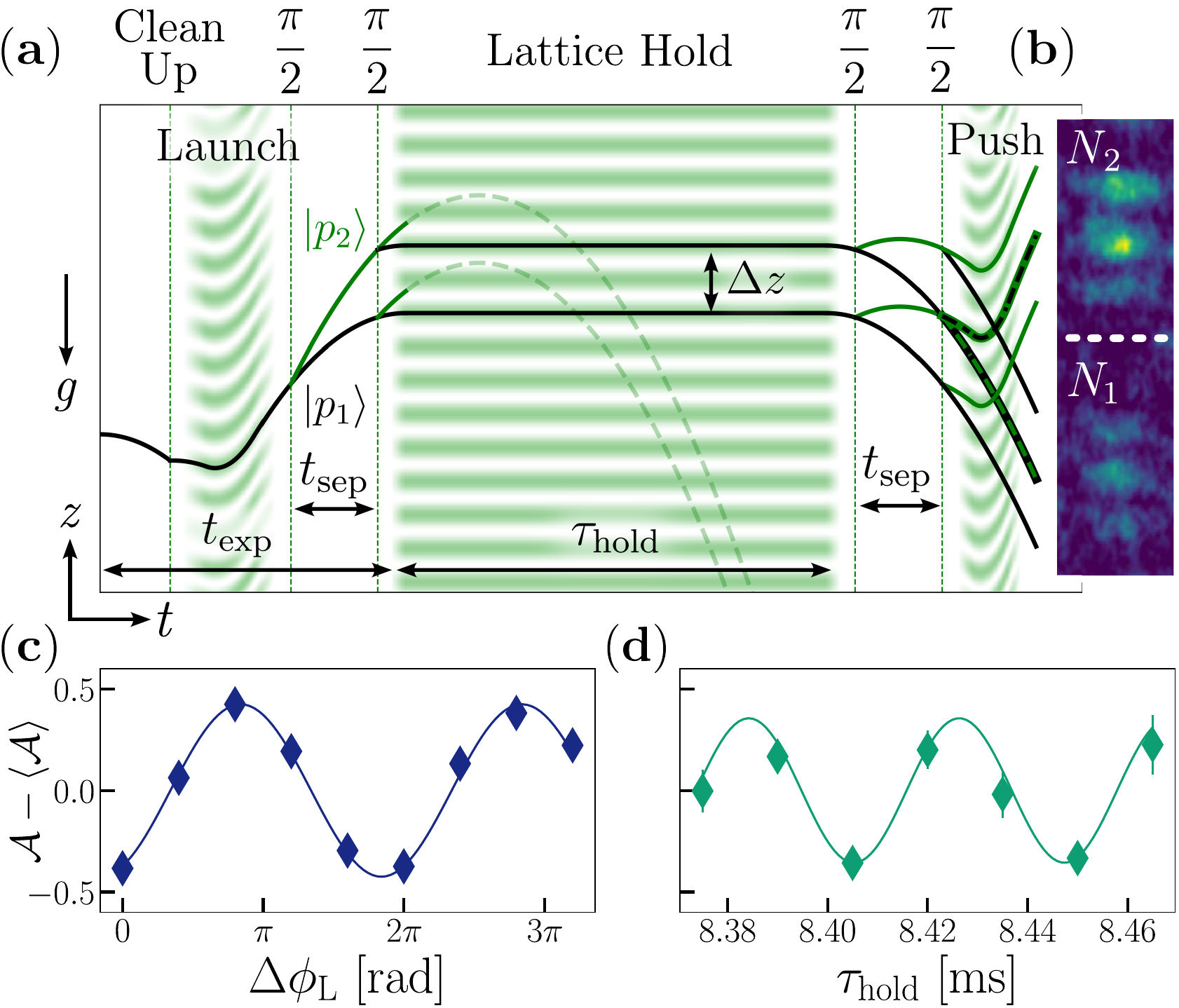}
     \caption{(a) Space-time diagram of the interferometer sequence. The BEC expands before velocity-selection with a first-order Bragg pulse (vertical dashed line ``Clean Up") and launch via Bloch oscillations (striped curves). Two $\pi/2$ pulses create the superposition loaded into the lattice trap at the kinematic apex. A symmetric pair of $\pi/2$ pulses close the interferometer and another BO sequence (``Push") spatially separates the output ports. (b) Representative absorption image at the interferometer output with all six distinct paths resolved (here $\Delta z = 120 d$). Representative interference fringes observed in the asymmetry, obtained by (c) varying the phase of the final $\pi/2$ pulse and by (d) varying the hold time. The frequency of (d) is proportional to $g$ and the wavepacket separation (here $\Delta z = 20 d$). Error bars represent the standard error on the mean.}
     \label{fig:schematic}
 \end{figure}

As shown in the schematic space-time diagram in Fig.~\ref{fig:schematic}(a), our interferometer is oriented vertically and is initiated by releasing the BEC from initial confinement at time $t=0$. The BEC then expands and falls under gravity before being coherently manipulated by standing wave atom optics. Pulsed optical standing waves coherently split the expanding source into two paths which are then confined and held with controlled separation for a variable interrogation time $\tau_{\rm hold}$ in an optical lattice trap. Upon release from this trap, further standing wave pulses coherently recombine these paths forming two output ports. The atom numbers $N_1$ and $N_2$ in the two ports oscillate with number asymmetry $\mathcal{A} \equiv (N_1-N_2)/(N_1+N_2)=C\cos(\Delta\phi)$ where $C$ is the interferometer contrast and
\begin{equation}
\label{eq:phi_prop}
\begin{split}
      \Delta \phi = 2\pi\left( \frac{\Delta z}{d} \right)\left( \frac{\tau_\mathrm{hold}+T'}{T_\mathrm{BO,g}} \right) + \Delta\phi_L
\end{split}
\end{equation}
is the interferometer phase accrued for a path separation $\Delta z$ during $\tau_{\rm hold}$. Here $d=\lambda/2$ is the lattice spacing for optical wavelength $\lambda$, $\tau_\mathrm{hold}+T'$ is the total time between the second and third $\pi/2$ pulses (Fig.~\ref{fig:schematic}), $\Delta\phi_L$ is a controllable laser offset phase (applied on the fourth $\pi/2$ pulse), and $T_\mathrm{BO,g} = \frac{h}{mgd}$ is the Bloch oscillation (BO) period for an atom with mass $m$ under gravitational acceleration $g$. Representative interferometer fringes obtained by varying either $\Delta \phi_L$ or $\tau_{\rm hold}$ are shown in Fig. \ref{fig:schematic}(c,d).

For a weakly-interacting BEC source, the expectation for non-interacting particles in Eq. (\ref{eq:phi_prop}) will be modified according to the Gross-Pitaevskii equation
\begin{equation} \label{eq:GPE}
\left[ -\frac{\hbar^2}{2m} \nabla^2+V(\vec{r})+\frac{4\pi\hbar^2a}{m}N|\psi(\vec{r})|^2 \right] \psi(\vec{r}) = \mu \psi(\vec{r}),
\end{equation}
which governs the evolution of the BEC wavefunction $\psi$ for $na^3 \ll 1$, where $a$ is the $s$-wave scattering length, $\mu$ is the chemical potential, $N$ is the total atom number, and $n=N|\psi(\vec{r})|^2$ is the number density. 

Interactions affect Eq.~(\ref{eq:phi_prop}) as a position dependence on $\Delta \phi$ which in turn affects the interferometer contrast. During BEC expansion, the interaction energy is converted into kinetic energy as the atomic density decreases. Capture and compression in the optical lattice trap leads to a subsequent density increase.

Our experimental controls provide a large dynamic range to investigate the effects of interactions associated with the narrow momentum distribution of the BEC source in lattice-trapped interferometers. Tuning the expansion time provides the tool to optimize source brightness while suppressing interaction effects.

\section{Experimental Setup \label{sec:exp_setup}}

\subsection{Atom Source and Optics\label{subsec:atom_source}}

Our experiment utilizes an apparatus and methods described in earlier work \cite{plot18,plot18thesis,goch19,mcal20,goch21,rahm24}. Briefly, following laser cooling of $^{174}$Yb, we load the atoms into a $532\,$nm crossed optical dipole trap (ODT). Forced evaporative cooling over $4\,$s produces pure BECs of $5 \times 10^4$ atoms with final ODT trap frequencies $\omega_i/2\pi=\{ 230, 43, 194\}\,\mathrm{Hz}$.  For these experimental parameters and $a\!=\!5.6\,$nm for $^{174}$Yb, we are in the Thomas-Fermi (TF) regime where the kinetic energy term in Eq.(\ref{eq:GPE}) is negligible. The corresponding peak density is $n_0\!=\!2.5 \times 10^{14} \ \mathrm{cm}^{-3}$, peak interaction energy is $\mu\!=\!h\times2.1\,$kHz $=k_B\times105\,$nK, and TF radii are $R_{0,i}=\sqrt{2 \mu/{m\omega_i^2}}=\{80,185,87\}\,\mu$m.

The atom optics are formed from two counter-propagating laser beams with $1/e^2$ radius (waist) $w_0\!=\!550\,\mu$m and red-detuned from the ${^1S}_0\leftrightarrow{^{3}P}_1$ intercombination line at $\lambda\!=\!556\,$nm with linewidth $\Gamma\!=\!2\pi\times182\,$kHz. The corresponding recoil energy is $E_\mathrm{rec} = \hbar^2 k^2/2m=h \times 3.71\,$kHz$=k_B\times185\,$nK. For all the experiments described below, the detuning $\Delta = -2\pi\times67\,\mathrm{GHz}=-3.6\times10^5\,\Gamma$ is chosen to minimize spontaneous scattering while providing enough lattice depth for the required atom optics given the available experimental laser power of up to $80\,$mW per beam. The frequency difference $\delta (t)$ between the beams is controlled at the sub-$\mathrm{Hz}$ level by two acousto-optical modulators driven by direct digital synthesizers which are frequency-locked to a $\mathrm{Cs}$ reference. The beams are aligned to local gravity to better than $1~\mathrm{mrad}$ via retro-reflection off a cup of water. A constant frequency chirp $\dot{\delta}_g=2kg\simeq2\pi\times35.2~ \mathrm{kHz/ms}$, where $k=\pi/d$ is the lattice wavevector, is applied through the entire interferometer sequence to compensate for the changing Doppler shift of the freely-falling atoms, with the exception of the trapping phase when $\delta=\dot{\delta}=0$.

\subsection{Interferometer Operation\label{subsec:ifm_op}}

An initial free expansion from the ODT for 3 ms reduces the density and interaction effects such that the velocity width of the expanding condensate is within 4\% of the asymptotic value \cite{supp}. Next, a velocity selection (``Clean Up") 1$^{\rm st}$-order Bragg pulse with a full width at half-maximum (FWHM) of $63 ~ \mathrm{\mu s}$ and peak Rabi frequency $2\pi\!\times\!1.7\,$kHz is applied. This reduces the axial velocity width of the ensemble to $0.1\hbar k$ and ensures the entire ensemble is addressed by the $\pi/2$ pulses, which have a FWHM of $32 ~ \mathrm{\mu s}$ and peak Rabi frequency $2\pi\!\times\!2\,$kHz. The beamsplitters couple states $\ket{p_1}$ and $\ket{p_2=p_1+2\hbar k}$. The total atom number entering the lattice trap can be controlled by tuning the peak Rabi frequency of the velocity selection pulse. The relative number between the two interferometer arms can be controlled by tuning the peak Rabi frequency of the first beamsplitter.

The BO-driven acceleration in the launch phase is accomplished by adiabatically loading into the ground band of an optical lattice with peak depth $U_0=5E_\mathrm{rec}$ over $400\,\mathrm{\mu s}$, then adding a relative frequency chirp of $2\pi\!\times100\mathrm{kHz/ms}$ to $\dot{\delta}_g$. Landau-Zener tunneling losses are negligible for these parameters \cite{rahm24}. A variable number of BOs between 7 and 30 set the range of total expansion time for the BEC before the lattice trap to between $6.8\,$ms and $26.2\,$ms. This changes the initial size, and thus density, of the ensemble in the lattice trap.

The separation between the wavepackets is set by the time $t_\mathrm{sep} $ between the two splitting $\pi/2$ pulses to $\Delta z\!=\!2 v_\mathrm{rec}t_\mathrm{sep}$, where $v_\mathrm{rec}\!=\!\hbar k /m\!=\!4.1 \mathrm{\mu m / ms}$. At the classical apex of the atomic trajectories, $\delta$ is set to zero and the lattice trap is adiabatically ramped to $U_0=4E_\mathrm{rec}$. The lower $\ket{p_1}$ momentum pair of the ensemble is thus loaded into the ground band of the stationary lattice trap, centered around quasimomentum $q\!=\!0$ with width $0.1\hbar k$. The higher $\ket{p_2}$ momentum pair loads into the first excited band and Landau-Zener tunnels, falling out of the trap \cite{supp}. During the lattice hold, the atoms Bloch oscillate due to gravity with period $T_\mathrm{BO,g}=836\,\mu$s. 

The atomic trajectories are additionally controlled by BO-driven path-selective acceleration during the ``push" phase to sufficiently separate the output ports for optimized detection \cite{picc22}. The push uses a constant number of BOs (usually 10) with the same lattice depth and sweep rate as the launch to separate the output ports by $\gtrsim$ 250 $\mathrm{\mu m}$ after a further $2\,$ms of time of flight before imaging.

\subsection{Signal Detection\label{subsec:sig_det}}

After recombination of the wavepackets, we measure the number of atoms in each spatially-resolved momentum state $N_1, \ N_2$ using standard absorption imaging on the singlet ${^1S}_0\leftrightarrow{^{1}P}_1$ transition at $399\,$nm (Fig.~\ref{fig:schematic}(b)). The three distinct paths in each port are usually overlapped and are only resolvable for small $t_\mathrm{exp}$ and large $\Delta z$. When investigating the contrast of the interferometer, we reproducibly change the laser phase by driving a frequency difference $\delta f_\mathrm{offset}$ for a variable time $t_\mathrm{ramp}$ between the third and fourth beamsplitters, producing a known phase shift $\Delta \phi_L=2\pi f_\mathrm{offset} t_\mathrm{ramp}$. In general, the total laser phase includes a small shift from imperfect cancellation of the gravitationally-induced Doppler shift: $(2kg-\dot{\delta}_g)t_\mathrm{sep}(t_\mathrm{sep}+T')\simeq0$. Typical parameters for the phase ramp are $10\ \mathrm{kHz}$ for $2\pi$ phase in 0.1 ms. We then fit the asymmetry data to $\mathcal{A} = \langle \mathcal{A} \rangle + C \cos(\phi_0+\Delta \phi_L)$, with $C$, phase offset $\phi_0$, and fringe offset $\langle \mathcal{A}\rangle$ as free parameters (Fig.~\ref{fig:schematic}(c)). The contrast is theoretically limited to 0.5 as only two of the four paths exiting the trap spatially overlap at the final beamsplitter. Alternatively, the interferometer hold time can be varied to obtain an interference fringe (Fig.~\ref{fig:schematic}(d)) with temporal frequency proportional to the local acceleration (Eq.~(\ref{eq:phi_prop})).

\begin{figure}
    \centering
    \includegraphics[width=\linewidth]{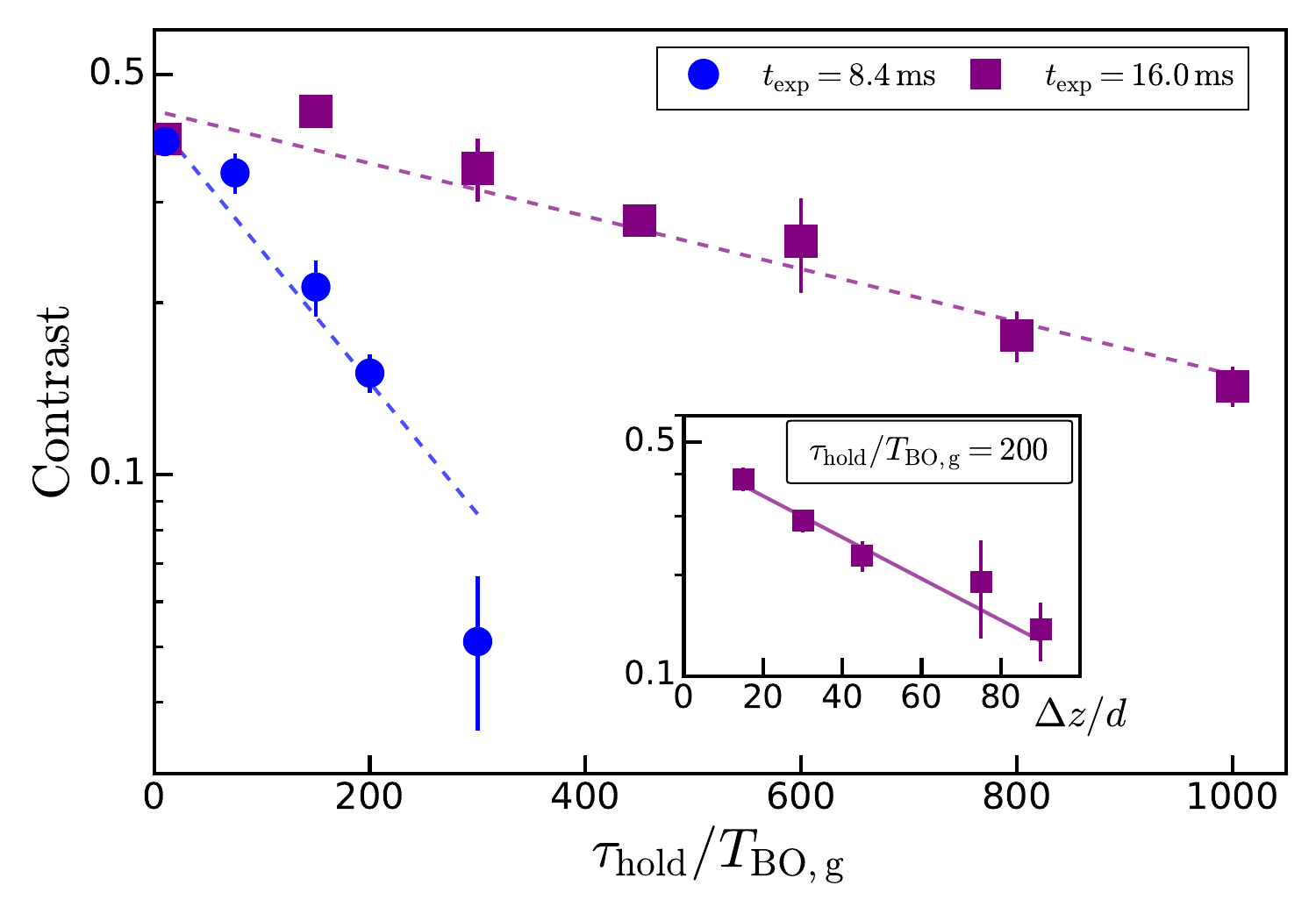}
    \caption{Contrast decay of the lattice-trapped interferometer with hold time for $\Delta z = 15d$ for BEC sources at two different expansion times. Dashed lines are exponential fits, with $1/e$ decay times $\tau_C/T_\mathrm{BO,g}=189(26),\ 941(113)$ for $t_\mathrm{exp}=8.4\,\mathrm{ms}, \ 16.0\,\mathrm{ms}$ respectively. Inset: contrast decay with wavepacket separation for a BEC source at $\tau_{\rm hold}=200\,T_\mathrm{BO,g}$ and $t_{\rm exp}=16.0\,$ms. Solid line is an exponential fit with $1/e$ decay length $\kappa/d=70(8)$. Error bars represent $1\sigma$ fit uncertainties.}
    \label{fig:C_vs_thold}
\end{figure}

\begin{figure*}
    \centering
    \includegraphics[width=\linewidth]{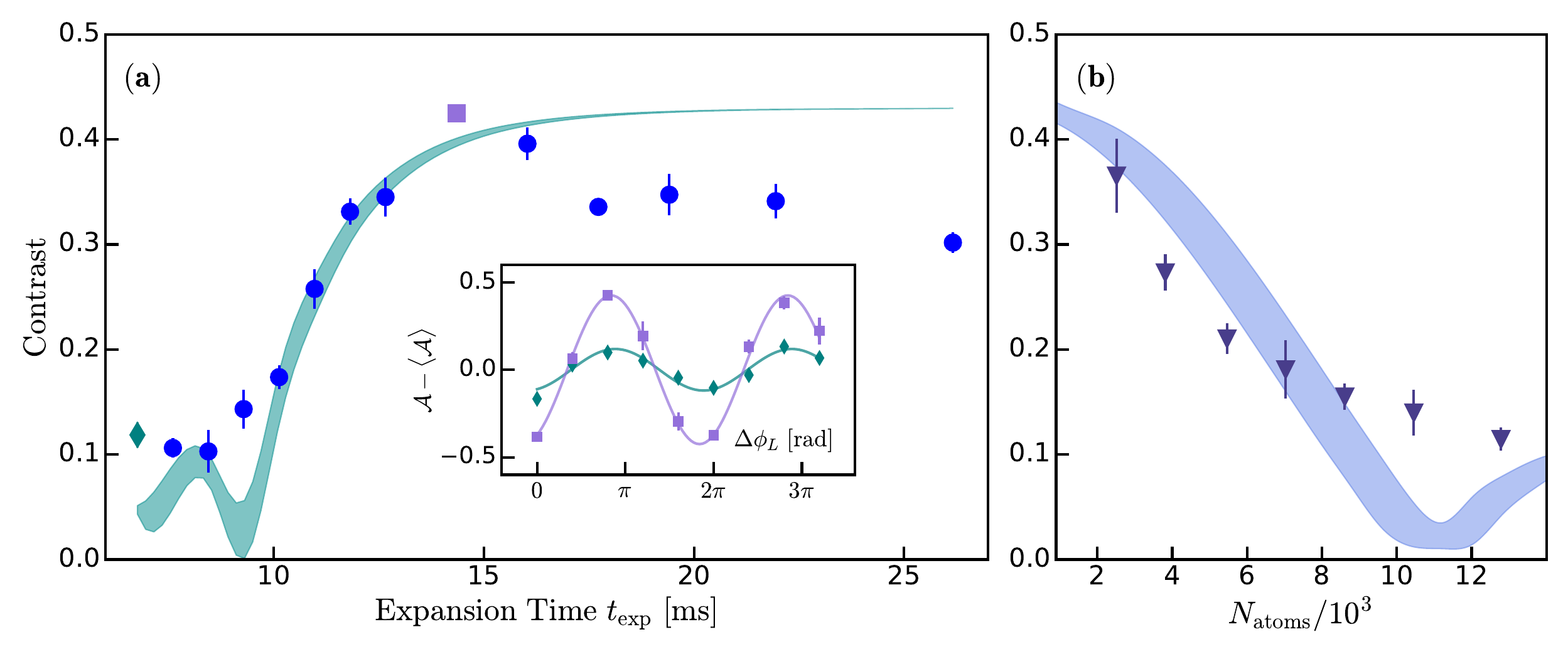}
    \caption{Control of the contrast at $\tau_\mathrm{hold}/T_\mathrm{BO,g}\!=\!200$ and $\Delta z / d\!=\!15$ by varying the density of the BEC source. (a) Observed contrast for various $t_{\rm exp}$. Inset: fringes at $t_{\rm exp}=6.8\,$ms (diamonds) and $t_{\rm exp}=14.5\,$ms (squares). (b) Observed contrast at $t_\mathrm{exp}=8.4\,$ms for various atom numbers. All error bars represent the standard error on the mean. Shaded bands in (a) and (b) show the theoretical results from a mean-field model incorporating typical 5\% number fluctuations (see text).}
    \label{fig:C_vs_density}
\end{figure*}

\section{Observation and Control of Interactions \label{sec:MFD}}

The use of a BEC source with adjustable expansion time gives access to a large dynamic range of atomic density and consequently interaction strength. As shown in Fig. \ref{fig:C_vs_thold}, changing the expansion time from $8.4\,$ms to $16\,$ms has a dramatic effect on the interferometer contrast. The observed $1/e$ coherence time $\tau_C$ is increased by a factor of five for a fixed wavepacket separation of $\Delta z = 15d$. 

The mean-field potential felt by the atoms during the lattice hold affects the performance of the interferometer in two distinct ways. First, the spatially-inhomogeneous density profile gives rise to a mean-field phase that varies across the ensemble and degrades the contrast of the interferometer output (Fig.~\ref{fig:C_vs_thold}). Second, any density imbalance between the two arms produces a global phase shift compared to the single-particle prediction of Eq.(\ref{eq:phi_prop}). We investigate both of these effects using three distinct experimental controls: the total expansion time of the BEC source, the total atom number entering the interferometric sequence, and the number imbalance between the two arms. These experiments are discussed in Secs. IVA and IVB, and the corresponding theoretical model and its scope are discussed in Secs. IVC and IVD. 

\subsection{Contrast Reduction and Recovery\label{subsec:contrast_reduction}}

A balanced interferometer with equal densities in the two arms will not experience any global phase shift due to mean-field interactions, but interactions can affect the contrast. We explore this systematically by controlling the BEC density using two independent methods: sample volume and total atom number. Both sets of experiments are performed at $\tau_\mathrm{hold}/T_\mathrm{BO,g}=200$ and $\Delta z / d =15$.

In a first set of experiments, we vary the number of BOs for the launch, which changes $t_{\rm exp}$ and the sample size in the lattice trap. We observe an increase in contrast by a factor of four (Fig.~\ref{fig:C_vs_density}(a)) with increasing $t_{\rm exp}$. An ensemble-averaged mean-field dephasing (MFD) model (shaded band, Sec. \ref{subsec:MFD}) captures this initial increase well. The width of the theory band corresponds to the typical experimental atom number fluctuations of 5\%. For larger $t_{\rm exp}$, the contrast decreases from a peak of approximately 0.4 at $t_\mathrm{exp}=14.5\ \mathrm{ms}$ to 0.3, and is discussed further in Sec. \ref{sec:decoherence}.

In a second set of experiments, we change the number of atoms entering the interferometer sequence by decreasing the peak Rabi frequency of the velocity selection pulse at fixed $t_\mathrm{exp}=8.4\,$ms. This isolates the effect of density at fixed volume and fixed velocity distribution. As shown in Fig.~\ref{fig:C_vs_density}(b), we see a recovery in the contrast with decreasing density and a good qualitative agreement with the MFD model. 

\subsection{Phase Shift\label{subsec:phase_shift}}

\begin{figure}[t]
    \centering
    \includegraphics[width=\linewidth]{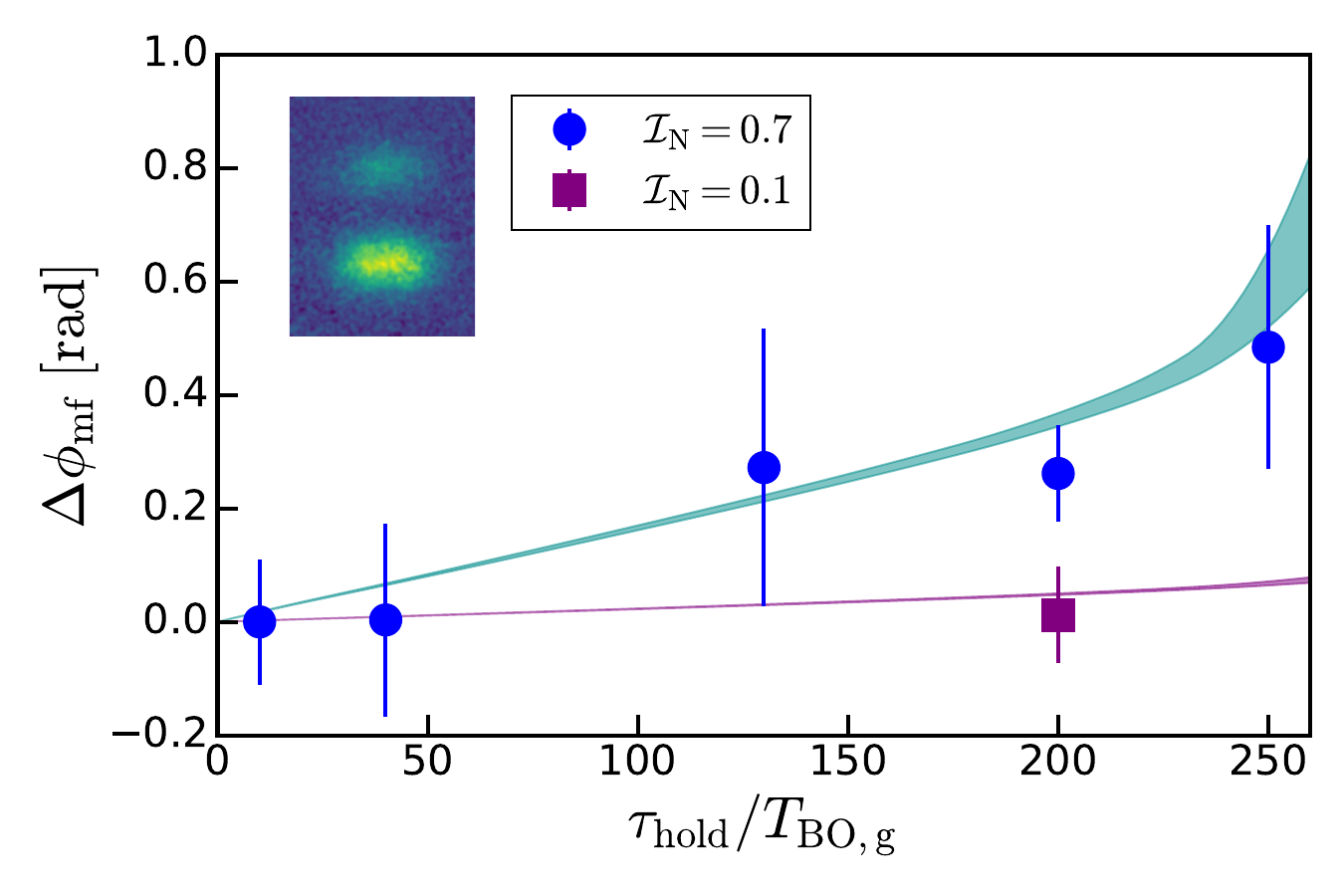}
    \caption{Global phase shift due to mean-field interactions observed by biasing the interferometer arms with an imbalance $\mathcal{I}_N$ of 0.1 (squares) and 0.7 (circles), for $t_\mathrm{exp}=8.4\,$ms. Error bars represent the standard error on the mean. Shaded bands show the theoretical results from a mean-field model incorporating typical 5\% number fluctuations (see text). Inset: absorption image of atoms after the first $\pi/2$ pulse showing a number imbalance of 0.7.}
    \label{fig:num_asym}
\end{figure}

In addition to dephasing the ensemble, mean-field interactions will also produce global phase shifts when there is a non-zero difference in the peak densities between the interferometer arms. We verify this by intentionally biasing the arms of the interferometer with different amplitudes and observing a global phase shift $\Delta \phi_\mathrm{mf}$. We create an arm imbalance $\mathcal{I}_N$, defined as the normalized difference between the atom number in the top and bottom arms entering the lattice trap, by reducing the peak Rabi frequency of the first beamsplitter and find reasonable agreement of the MFD model with our data (see Fig.~\ref{fig:num_asym}) with no free parameters. 

These data were taken by comparing the phase $\phi_0$ of a fringe obtained by scanning the phase of the final beamsplitter using a BEC source at fixed expansion time ($t_\mathrm{exp}=8.45\mathrm{ms}$) and a thermal source with an identical interferometric sequence. The large imbalance of 0.7 results in a low contrast of about 0.1 for the corresponding fringes, leading to somewhat large error bars and limited interrogation time ($\lesssim250\ T_\mathrm{BO,g}$) in Fig.~\ref{fig:num_asym}.

\subsection{Theoretical Model\label{subsec:MFD}}

We model our system using the Castin-Dum treatment \cite{cast96} of interacting BECs in the Thomas-Fermi regime during the expansion time $t_{\rm exp}$ until lattice trapping. After release from the ODT, the BEC wavefunction undergoes a self-similar dilation \cite{jami11} with TF radii growing as $R_i(t)=\lambda_i(t)R_{0,i}$. We determine the scale factors $\lambda_i (t)$ by numerical integration \cite{supp} and obtain the condensate density during expansion as $n(\vec{r},t) = n_0(t)(1-\sum_i r_i^2/R_i(t)^2)$ when $n(\vec{r},t) \geq 0$ and equal to zero otherwise. Here $n_0(t)=n_0(0) \prod_i 1/\lambda_i(t)$ is the peak condensate density. 

Incorporating the effects of velocity selection on the number and velocity width in the axial dimension provides reconstruction of the full three-dimensional density profile at the time of trapping $t_\mathrm{exp}$ \cite{supp}. We assume the Castin-Dum solution is unaffected by the launch, which occurs once the expansion has transitioned from hydrodynamic to ballistic. The adiabaticity in the axial dimension of the BO process preserves the axial density during the launch, and the total launch time is much shorter than the transverse oscillation period leading to negligible corrections to the transverse density.

Once adiabatically loaded into the ground Bloch state, the lattice-trapped ensemble is not a BEC, but can be treated as a quasi-2D bosonic gas at each occupied lattice site with effective mean-field coupling $8\pi\hbar^2a\xi/m$, where
\begin{equation} \label{eq:g2d}
    \xi = \frac{1}{2\hbar k}\int^{\hbar k}_{-\hbar k} dq \ \int_0^d dz \ \left| \Psi_q(z)\right|^4
\end{equation}
is the average in both position over one lattice site and quasimomentum over the Brillouin zone of the ground state Bloch wavefunction $\Psi_q(z)$, obtained by numerical diagonalization of the lattice Hamiltonian at $U_0=4E_\mathrm{rec}$. This gives rise to a spatially-varying phase 

\begin{equation} \label{eq:phi_mf}
    \varphi_\mathrm{mf}(\vec{r})=\frac{8\pi\hbar a\xi}{m} n(x,y,z_i;t_\mathrm{exp})\tau_\mathrm{hold}
\end{equation}
where $z_i$ label the positions of the lattice sites occupied by the ensemble. The number of atoms at each site is computed by integrating over the axial density. We neglect the dynamics of the density profiles in a static approximation, i.e. the transverse density is assumed to not change after loading into the trap. This inhomogeneous mean-field phase changes the interferometric output 
\begin{equation}
\begin{split}
    \mathcal{A}\rightarrow\langle \mathcal{A}(\vec{r}) \rangle &= C \ \Re \left[ e^ {i \Delta \phi} \langle e^{i \Delta \varphi_\mathrm{mf}(\vec{r})}  \rangle \right] \\\\ &= C  |\eta| \cos (\Delta \phi + \mathrm{Arg[\eta]}),
\end{split}
\end{equation}
which reduces the contrast by a factor of $|\eta|$, where $\eta\equiv \langle e^{i \Delta\varphi_\mathrm{mf}(\vec{r})} \rangle$ and brackets denote ensemble averages in both the transverse and axial dimensions. This mean-field ensemble dephasing model has good qualitative agreement with both the expansion timescale and number scale at which the observed contrast at $200 \ \mathrm{BOs}$ changes in both experiments shown in Fig \ref{fig:C_vs_density}, where the theory bands are $C_0|\eta|$ with $C_0=0.43$ to reflect the maximum observed value at $t_\mathrm{exp}=14.5\,\mathrm{ms}$. In a biased configuration, the global phase shift $\Delta \phi_\mathrm{mf}$ is given by $\mathrm{Arg} [\eta]$. The differential mean-field phase $\Delta\varphi_\mathrm{mf}(\vec{r})$ is computed by taking the site-wise difference between sites at positions $z_i$ and $z_i+\Delta z$, including the axial overlap of the upper and lower arms \cite{supp}.

\subsection{Scope of Model \label{subsec:Scope}}

During the lattice hold time, the ensemble is no longer a BEC and is also not in thermal equilibrium since the peak collision rate $\Gamma_\mathrm{col}=n_0 (8\pi a^2) v_\mathrm{rms}$ is less than $1\,$Hz and negligible for all parameters investigated in this work. This makes the transverse dynamics challenging to model. In a harmonic approximation, the transverse oscillator frequency is $\omega_\perp\!=\!\sqrt{4U_0/(mw_0^2)}=2\pi\times3.4\,\mathrm{Hz}$, and the corresponding breathing mode period is $150 \ \mathrm{ms}$. Our observations at $\tau_{\rm hold}=200 \ T_\mathrm{BO,g}=170 \ \mathrm{ms}$ (Fig.\ref{fig:C_vs_density}) are still well-captured by our static model, suggesting that any transverse oscillations in cloud size effectively averages away with minimal effect on the interferometer signal.    

The model underestimates the contrast in regimes of high density (Fig.\ref{fig:C_vs_density}) where mean-field effects are the strongest. This is consistent with the expectation from a static model which ignores interaction-driven dynamics, as the substantial phase gradients responsible for such large dephasing rates in this regime give rise to velocity fields that act to decrease the density and dephasing rate. 

This model also overestimates the contrast at the largest expansion times in Fig. \ref{fig:C_vs_density}(a) when the cloud size is largest and interaction effects are negligible. The contrast maximum around $t_\mathrm{exp}=14.5\,\mathrm{ms}$ could be related to the length scale of lattice beam imperfections present in our system or to spatial mode-matching of the atom cloud with lattice beam size. 

\section{Interferometer Performance} \label{sec:IfmPerf}

The use of a BEC source has allowed us to effectively probe the effects of atomic interactions in lattice-trapped interferometers. With an understanding of how to control these effects, we now turn to the metrological performance of our interferometer. In this Section, we first present our findings on how the brightness of a BEC source allows it to outperform a thermal atom source, followed by a discussion of the current limiting decoherence mechanisms and an assessment of the interferometer sensitivity for accelerometry and gravimetry. 

\begin{figure}
    \centering
    \includegraphics[width=\linewidth]{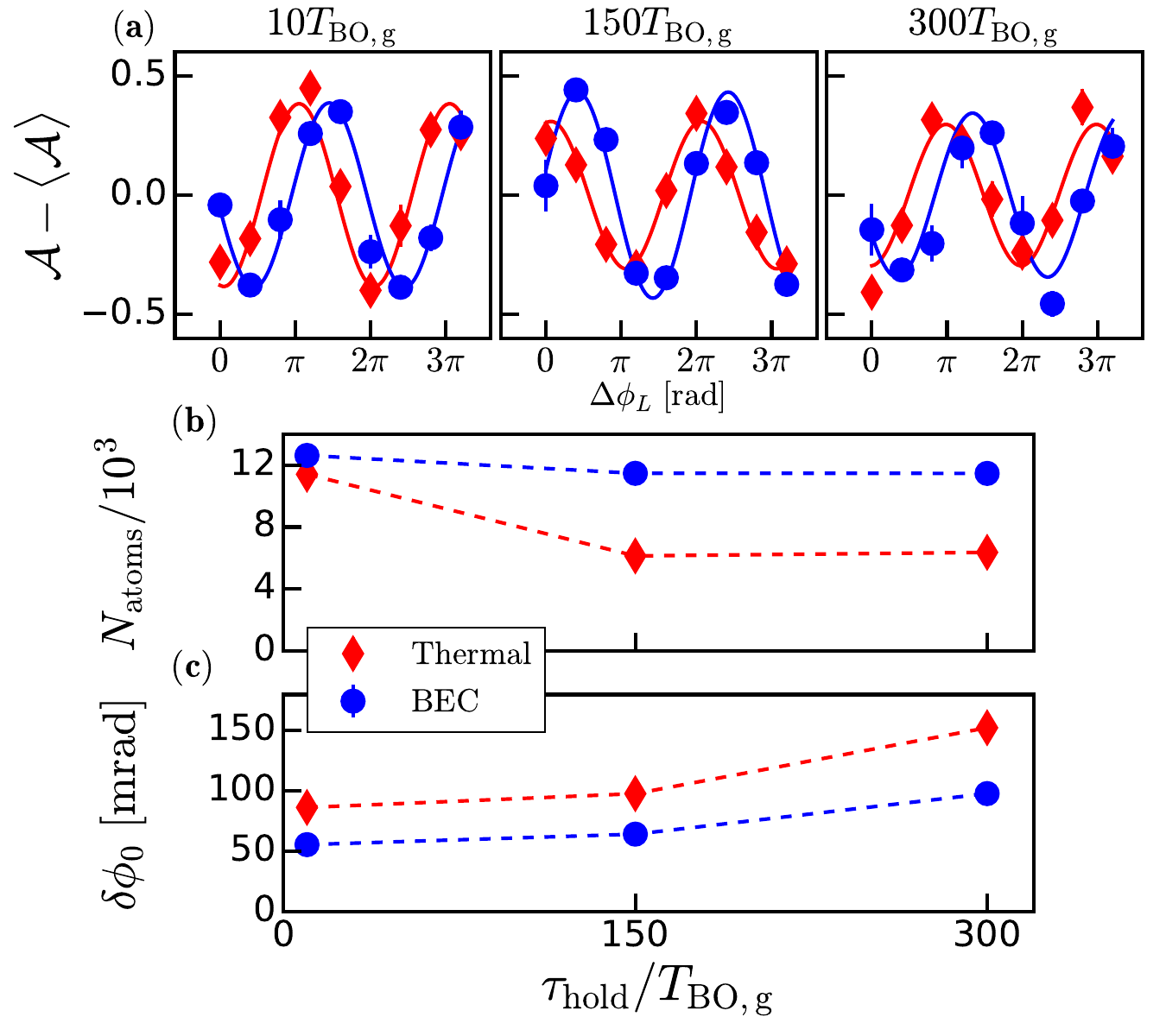}
    \caption{Comparison of BEC ($t_\mathrm{exp}=16\,$ms) fringes with those from a thermal source (initially at $350\,$nK, $t_\mathrm{exp}=8.4\,$ms). (a) Fringes at three hold times for $\Delta z = 15d$ for thermal (red diamonds) and BEC (blue circles) sources. (b) Total atom number as a function of hold time. (c) Statistical phase error from sinusoidal fits. The thermal data has an average of 5 points per phase scanned, whereas the BEC has 4.}
    \label{fig:therm}
\end{figure}

\subsection{BEC vs Thermal}

We compare the performance of a BEC source to that of a thermal source by directly comparing the contrast and phase error of the corresponding interference fringes. For the thermal source, we initiate the interferometry sequence after stopping the evaporative cooling sequence at $350\,$nK, before condensation occurs in the ODT. Fringes, atom numbers, and phase errors are shown for three values of $\tau_{\rm hold}$ for the BEC and thermal cloud cases in Fig.~\ref{fig:therm}.

We observe fringe contrasts that are comparable for the two cases (Fig.~\ref{fig:therm}(a)). However, for the thermal source, the larger initial size and transverse velocity distribution lead to significant atom loss within a single transverse oscillation period through spilling and tunneling processes (Fig.~\ref{fig:therm}(b)). We observe a factor of two reduction in the total atom number compared to the BEC source \cite{footnum}. This is consistent with the $4E_\mathrm{rec}\simeq750~\mathrm{nK}$ trap depth. This process also narrows the transverse velocity distribution and the subsequent velocity width due to position-velocity correlations. The BEC thus provides a larger steady-state atom flux compared to a thermal source, consistent with its higher phase space density. 

As shown in Fig.~\ref{fig:therm}(c)), we also observe a larger statistical phase error in the thermal case (up to a factor of 2), despite the $25\%$ greater number of experimental shots compared to the BEC case. We attribute this to ultimately stem from the larger atom number or brightness of the BEC source. 

We observe no statistically significant difference in the coherence times of close to 1\,sec  for the two sources in the regime not affected by mean-field effects (Fig.~\ref{fig:therm}(a) and \cite{supp}). Further exploitation of the BEC source beyond the advantages in atom number and phase error observed in our lattice-trapped interferometer will require assessing and removing the limiting decoherence mechanisms. 

\subsection{Decoherence Mechanisms \label{sec:decoherence}}

Increasing the coherence time $\tau_C$ and the coherence length $\kappa$ are key to scaling up the performance of the lattice-trapped interferometer. The spontaneous scattering rate at $\Delta/\Gamma\!=\!-3.6\times 10^5$ and $U_0\!=\!4E_r$ is $0.25\,$s$^{-1}$. Atoms which spontaneously scatter photons will remain in the trap until they acquire enough kinetic energy to leave, contributing to the decoherence rate $\tau_C^{-1}$. Indeed, we find that $\tau_C$ scales with $|\Delta|$ and $U_0$ in a manner consistent with spontaneous scattering. Imperfections in the lattice beam transverse profile could also contribute to the observed coherence time, but their analysis is beyond the scope of this work.  

Two effects that can contribute to the finite coherence length $\kappa$ (inset of Fig.~\ref{fig:C_vs_thold}) are imperfections in the lattice beam axial profile and phase noise from vibrations. Beam imperfections will increase the velocity spread of the atomic distribution exiting the interferometer. However, the magnitude of this spread is less than $0.01 v_\mathrm{rec}$, far below our detection threshold, even if we assume that $\kappa$ is limited entirely by this effect \cite{supp}. 

Phase noise from random vibrations can change the phase of the lattice between beamsplitters and scales with separation, reducing the visibility of the signal. $\Delta z = 100\,d$ requires $t_\mathrm{sep}=3.35\,$ms, which sets the frequency scale of $\sim300\mathrm{Hz}$. The spectral density of phase fluctuations above this value decrease in our setup, as measured with a heterodyne technique. This timescale could be improved in the future by passive or active \cite{hens99,lede25} vibration isolation or by using BO-enhanced large momentum transfer beamsplitters \cite{fitz24}.

\begin{figure}
    \centering
    \includegraphics[width=\linewidth]{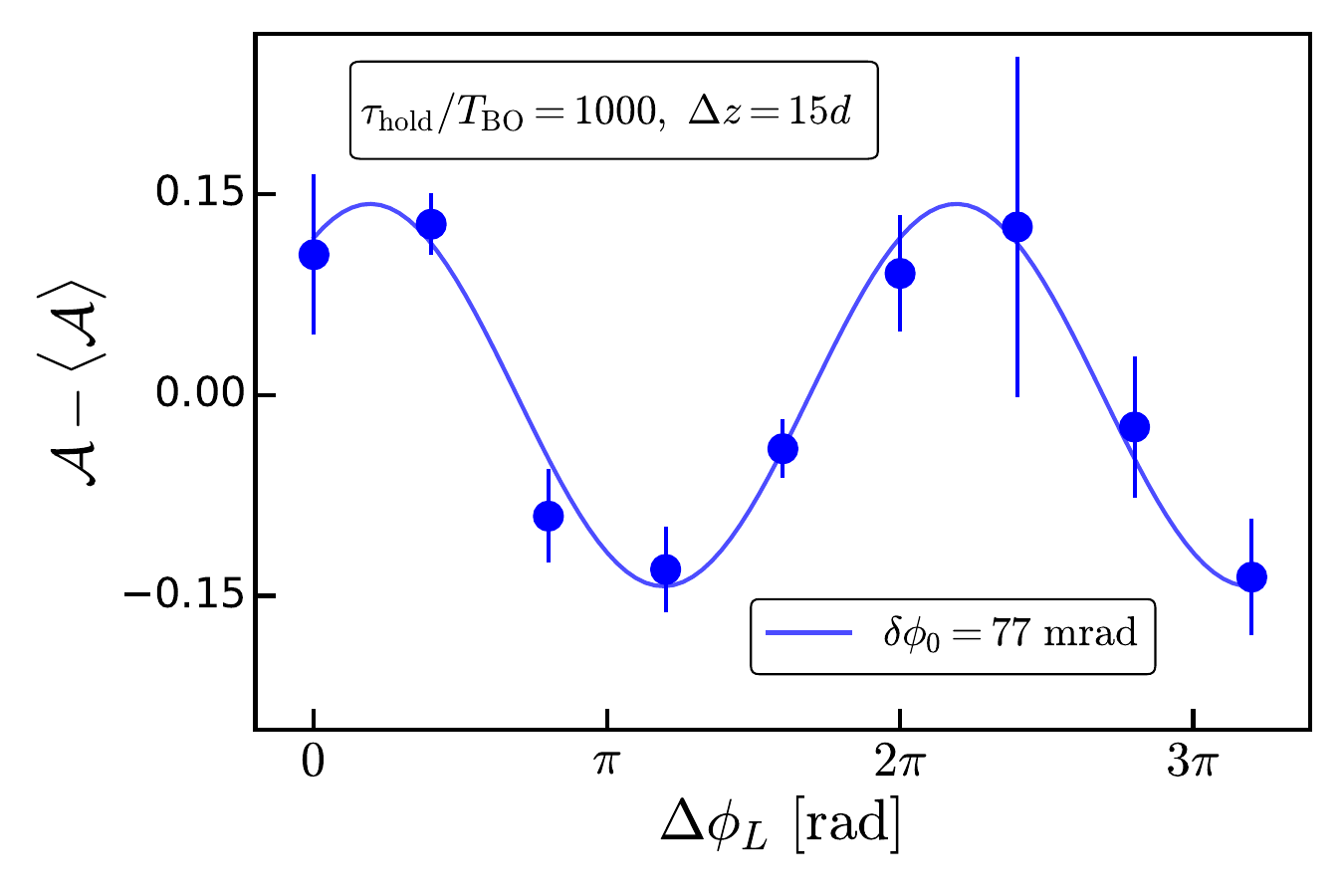}
    \caption{Fringe at $\tau_\mathrm{hold}=1000\,T_\mathrm{BO,g}$ and $\Delta z = 15\,d$. The $1\sigma$ statistical phase error of the fit is $\delta \phi_0=77\,$mrad. Error bars represent the standard error on the mean.}
    \label{fig:g}
\end{figure}

In the absence of beam imperfections such as in a cavity-based lattice, the BEC source should provide additional benefit from its reduced transverse velocity width which would reduce the dynamical phase spread from motion in the trap \cite{pand24}. The use of a cavity would also allow further detuning of the lattice trap and therefore reduction of spontaneous scattering for the same available laser power. 

\subsection{Acceleration Sensitivity}

Our interferometer is sensitive to the value of local $g$ (more generally, any inertial force) through the phase accumulated during the lattice trapping (Eq.~\ref{eq:phi_prop}). The measurement sensitivity is related to the phase uncertainty by $\delta g/g = \delta \phi/\phi$. For $\tau_{\rm hold}\!=\!1000\,T_\mathrm{BO,g}$ and $\Delta z\!=\!15\,d$ the total phase is approximately $10^5$ rad and the statistical uncertainty on the phase of the BEC source is $77\,$mrad (Fig.~\ref{fig:g}) for a total relative uncertainty of $0.8\times10^{-6}$. Alternatively, gravimetry and accelerometry can also be performed by monitoring the interferometer phase as a function of $\tau_\mathrm{hold}$ \cite{supp}.

\section{Conclusion and Outlook}

We have demonstrated the first Yb trapped atom interferometer, and introduced BEC sources to hybrid Ramsey-Bordé-Bloch lattice-trapped atom interferometry, a promising platform for metrology and sensing. We observe interaction effects on both interferometer contrast and phase in regimes of high BEC density. We demonstrate control over these effects through changing the total expansion time and atom number. We use a mean-field description of interactions to successfully model our experimental results. Within the explored parameter space, the BEC source is found to outperform a thermal source in both phase error and surviving atom number in the lattice hold.

The best coherence times of about one second observed in this work are limited by beam imperfections from free-space lattice optics and the spontaneous scattering rate which affect the BEC and thermal sources identically. The addition of a buildup cavity in the future would mitigate both decoherence sources and allow further extraction of advantage from the use of a near-monochromatic BEC source. While current acceleration sensitivity is still a few orders of magnitude below the state of the art \cite{zhan23}, future experimental upgrades could allow for competitive accelerometry, gravimetry, and an equivalence principle test using multiple Yb isotopes.

This work also demonstrates how lattice-trapped interferometers can help quantify weak interactions that manifest over hundreds of milliseconds. The observed interferometer phase shifts could be used to probe beyond mean-field effects and out-of-equilibrium dynamics in the lattice sites \cite{rey04,polk04}. The demonstrated interaction control by tuning the density could also be adapted towards implementing sub-shot noise interferometry using squeezed motional states \cite{sore01, corg21_2, corg21}.

\begin{acknowledgments}
We thank the group of H. Mueller [UC Berkeley], Daniel Gochnauer, and Daniel Allman for helpful discussions. We thank Richard Kim and Aidan Kemper for experimental assistance. This work was supported by ONR Grant N000142412564 and NSF Grant No. PHY-2110164. EH acknowledges support from the Department of Defense (DoD) through the National Defense Science \& Engineering (NDSEG) Fellowship Program. SG acknowledges support from the Job \& Gertrud Tamaki Chairship. 
\end{acknowledgments}


\begin{thebibliography}{45}
\bibitem [{foo()}]{footnum}%
  \BibitemOpen
  \bibfield  {title} {\bibinfo {title} {{When decreasing the velocity distribution from thermal to BEC, the two competing effects of narrowing velocity distribution and decreasing initial atom number roughly cancel after our velocity selection and approximately 25,000 atoms enter the interferometer sequence for both sources. The interferometer geometry (Fig. 1) implies that only half of these are trapped and contribute to the interferometer signal}}}\href@noop {} {\ }\BibitemShut {NoStop}%
\end{thebibliography}

%

\clearpage

\onecolumngrid

\renewcommand{\theequation}{S\arabic{equation}}
\renewcommand{\thefigure}{S\arabic{figure}}
\setcounter{equation}{0}
\setcounter{figure}{0}

\makeatletter

\newcommand*{\addFileDependency}[1]{
\typeout{(#1)}

\@addtofilelist{#1}
%
\IfFileExists{#1}{}{\typeout{No file #1.}}
}\makeatother

\newcommand*{\myexternaldocument}[1]{%
\externaldocument{#1}%
\addFileDependency{#1.tex}%
\addFileDependency{#1.aux}%
}

\renewcommand{\theequation}{S\arabic{equation}}
\renewcommand{\thefigure}{S\arabic{figure}}

\setcounter{MaxMatrixCols}{10}

\pdfoutput=1

\begin{center}
{\bf \large Supplementary Material for:\\ Lattice-Trapped Atom Interferometry with a Bose-Einstein condensate:\\ Observation and Control of Interactions}
\end{center}

\twocolumngrid

\section*{SI. Experimental Details}

The Bloch oscillation (BO) ``push" phase acts as a momentum-state-selective control of the spacetime trajectories of the atoms. Here we want to add momenta to the larger momentum state ($\ket{p_2}$) and leave the other state ($\ket{p_1}$) unaffected. To do so, we adiabatically load the $\ket{p_2}$ atoms into the ground band ($b=0$) of the optical lattice at an initial quasimomentum of $q_0=0.5\hbar k$, which loads the $\ket{p_1}$ atoms into the first excited band ($b=1$) at $q=0.5\hbar k$, away from the avoided crossings at $q_0=0$ and $\pm\hbar k$ to avoid mixing higher momentum states during the loading and BO processes. This is accomplished by loading with a frequency difference between the beams of $1.5\times14.85\,\mathrm{kHz}$ in the frame of the $\ket{p_1}$ atoms, where $4f_\mathrm{rec}=14.85\,\mathrm{kHz}$ is the Doppler shift associated with the Brillouin zone width of $2\hbar k$.

The Landau-Zener probability of tunneling during a Bloch oscillation can be expressed as 

\begin{equation}
    P_\mathrm{LZ} = \exp\left[-\frac{\pi \Omega_\mathrm{bg}^2}{2\beta\dot{\delta}}\right],
\end{equation}
where $\hbar \Omega_\mathrm{bg}$ is the band gap at the avoided crossing, $\beta$ is the higher of the two band numbers participating in the avoided crossing, and $\dot{\delta}$ is the rate of change of the angular frequency difference between the two beams \cite{mcal20, rahm24}. The band gap between bands 0 and 1 and between bands 1 and 2 is found by numerically diagonalizing the lattice Hamiltonian. For our sweep rate of $\dot{\delta}=\dot{\delta_g}+2\pi\times100\,$kHz/ms and a lattice depth of $5E_\mathrm{rec}$, the survival probabilities $(1-P_\mathrm{LZ})^{N_\mathrm{BO}}$ after $N_\mathrm{BO}=10$ BOs are 0.97 and $10^{-7}$ for the $\ket{p_2}$ atoms loaded into the ground band and the $\ket{p_1}$ atoms loaded into the excited band, respectively. This 3\% loss of $\ket{p_2}$ atoms artificially reduces the contrast for the push phase. The percent reduction in the contrast is equal to the percent of atoms lost in the push process, which at 3\% is less than the typical fit uncertainties of 10\%. More details of using BOs to physically separate the output ports of Bragg interferometers can be found in Ref.~\cite{picc22}.

During the load of the lattice trap, the $\ket{p_1}$ ($\ket{p_2}$) pair is loaded into the ground (first excited) band of the stationary optical lattice, respectively. At $U_0=4E_\mathrm{rec}$, the Landau-Zener tunneling probability between $b=0$ and $b=1$ is $3.5\times10^{-7}$ per BO, and 0.67 between $b=1$ and $b=2$. After 10 BOs, the number of atoms remaining in the excited band drops to 2\% and is negligible for longer hold times. To understand why the $\ket{p_2}$ pair tunnels out of the lattice trap, consider the momentum-space representation. These $\ket{p_2}$ atoms are loaded into the first excited band ($b=1$) at the beginning of the lattice trap and tunnel only to the second excited band ($b=2$) after a time of approximately half of $T_\mathrm{BO,g}$. After one full BO, gravity has reduced the velocity in the lab frame by exactly $2\hbar k$ (by definition), and thus the $\ket{p_2}$ atoms are stationary in the lab frame, but are still in excited Bloch states (either $b=1$ with a probability of of 0.33 of $b=2$ with a probability of 0.67) and are thus not trapped. In momentum space, after the classical apex in the lab frame, they continue to Landau-Zener tunnel to higher excited bands and fall out of the lattice trap. An alternative way of thinking about this process is as follows: the energy of atoms in the $\ket{p_2}$ state is modified by the adiabatic load into the first excited band, adding a node to the spatial wavefunction in the axial dimension. This additional energy is larger than the depth of the confining potential in the transverse dimension and is thus not a bound state.

\section*{SII. Density Model - State Preparation}

The theoretical mean-field model of our system is outlined in Sec. IV~C of the main text. Here we provide more detail about modeling the effects of velocity selection and the initialization of the density profiles in each lattice site of the optical lattice trap.

The effect of velocity selection on the axial density is modeled by numerically integrating the Schr\"odinger equation with a Bragg coupling in a plane wave basis for a range of relative velocities and against our specific pulse profile. This gives a diffraction probability for each velocity $v$, which is multiplied by the axial velocity distribution $n(v_z,t)\propto\left[ 1-v_z^2/v_{z,\mathrm{max}}^2(t)\right]^2$, where $v_{z,\mathrm{max}}(t) = R_z(t)\dot{\lambda}_z(t)/\lambda_z(t)$, of the BEC at the time of the pulse. Here $\lambda(t)$ are the Castin-Dum scale parameters and $R_z(t)$ is the condensate radius. 

\begin{figure}[t]
    \centering
    \includegraphics[width=\linewidth]{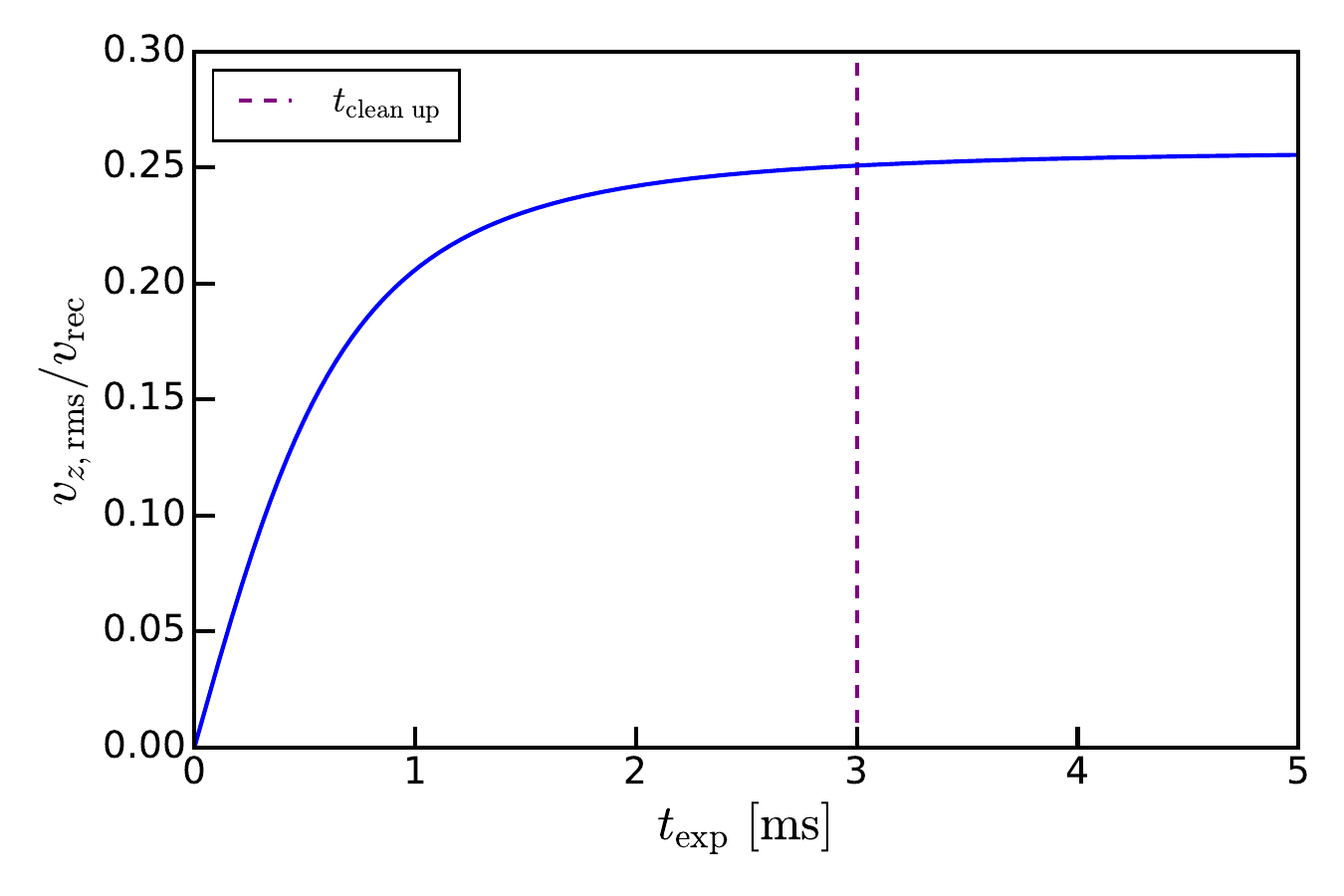}
    \caption{Calculated velocity width (root mean square) of the expanding BEC from the Castin-Dum solution. At the time of the velocity selection (``Clean Up") pulse, $t_\mathrm{clean\ up}=3\,\mathrm{ms}$, the axial velocity width has increased to within 4\% of the asympotoic value.}
    \label{fig:velocity_distribution}
\end{figure}

As mentioned in the main text, the hydrodynamic expansion has increased the velocity distribution to within 4\% of its asymptotic value at the time of the velocity selection pulse, as seen in Fig.~\ref{fig:velocity_distribution}, so we neglect any mean-field effects on the velocity distribution after the velocity selection pulse and treat the expansion in a self-similar ballistic manner. The peak mean-field energy is estimated from the Castin-Dum solution to be $h\times100\,\mathrm{Hz}$, representing a sub-percent shift of the Bragg resonance.

The hydrodynamic position-velocity correlation is maintained through the velocity selection, and thus the ballistic expansion in the axial dimension is defined by the local velocity field $v_z(z,t)=z\frac{\dot{\lambda}_z(t)}{\lambda_z(t)}$ for $z<R_z(t)$ and the time of flight from the end of the velocity selection to the time of trapping in the lattice, minus the launch time due to the adiabaticity of the load and launch in the axial dimension. The real-space axial density at the time of trapping $n_z(z,t_\mathrm{exp})$ is normalized to unity and the percent of the ensemble in each site of the lattice trap is calculated as an integral of $n_z(z,t_\mathrm{exp})$ over each lattice site. The axial density varies slowly on the scale of the lattice constant, so we approximate this integral as $N_\mathrm{atoms}(z_i)/N_\mathrm{atoms,tot}=d \times n_z(z_i,t_\mathrm{exp})$ where $d$ is the lattice constant and $z_i$ label the lattice sites.

The 2D transverse density at each lattice site is then constructed by the inverted parabola given by the Castin-Dum solution at time $t_\mathrm{exp}$, as in the main text, with condensate radii rescaled according to the axial position $z_i$. Each 2D density is normalized according to the number of atoms in that site $N_\mathrm{atoms}(z_i)$. From this initial density in each site, we calculate the mean-field phase according to the effective 2D coupling, giving a phase at each point in the 2D grid. For a given separation $\Delta z$, the differential mean-field phase $\Delta \varphi_\mathrm{mf}(x,y,z_i)$ is calculated between the site of interest at $z_i$ and that at $z_i + \Delta z$. The differential phase field is then ensemble-averaged in 2D with the initial transverse Thomas-Fermi distribution and over the axial dimension with the velocity-selected and ballistically-propagated axial distribution for the full 3D ensemble-averaged mean field phasor $\eta$.

\begin{figure}[t]
    \centering
    \includegraphics[width=\linewidth]{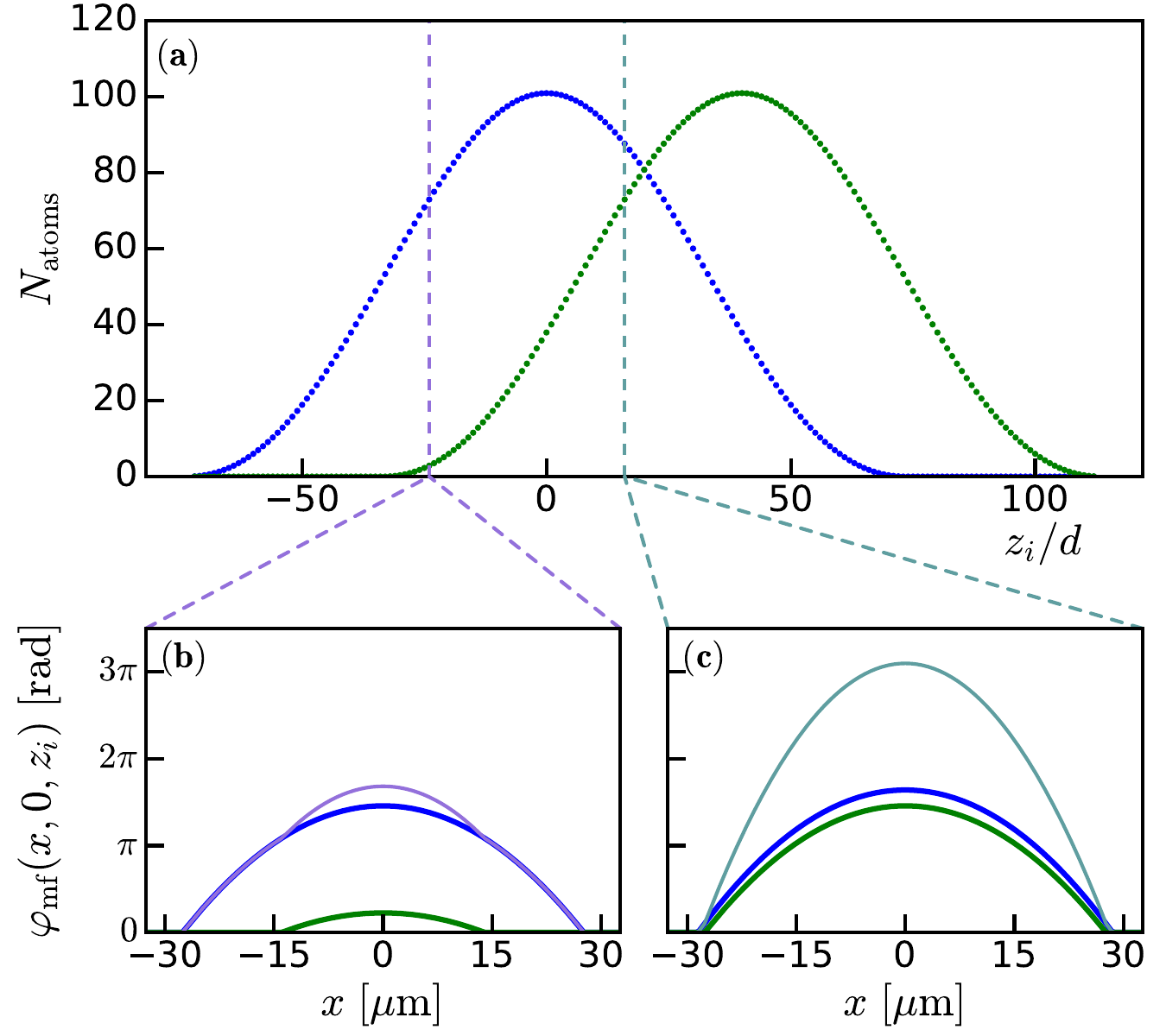}
    \caption{Overview of the differential mean-field phase with finite overlap of the two arms in the axial dimension. (a) Number of atoms in each lattice site from the bottom arm (blue, left) and top arm (green, right) with 14,000 total atoms, a separation of $\Delta z /d= 30$, and an expansion time $t_\mathrm{exp}=9\,$ms. (b,c) Transverse mean-field phase profiles along y=0 after a hold time of $\tau_\mathrm{hold}/T_\mathrm{BO,g}=200$. The lower lattice site (b) has a small enhancement of the mean-field phase from overlap with atoms in the upper arm (green), whereas the upper lattice site (c) has a large enhancement from overlap with atoms in the lower arm (blue).}
    \label{fig:overlap}
\end{figure}

Due to the finite spatial extent of each arm of the interferometer at the time of trapping, for all experimental configurations investigated in the main text there is non-negligible overlap of the atomic wavepackets from each arm in a given lattice site. This overlap leads to the spatially-inhomogeneous differential shift that gives rise to the reduction in contrast, as seen in Fig.~\ref{fig:overlap}. In the biased configuration, the effective imbalance between lattice sites is reduced by the overlap.

\section*{SIII. Lattice Imperfections}

As introduced in Section V~B of the main text, one source of decoherence is identified as imperfections in the axial mode of the optical lattice due to speckle or stray reflections. We can estimate the effect of such imperfections on the contrast as follows: the axial component of the velocity distribution acquires a random differential spread $\delta v$ from dipole forces arising from the imperfections. This $\delta v$ in turn introduces a random phase shift $\delta \phi = 2k\delta v t_\mathrm{sep}$ \cite{char12, zhan16}. This reduces the contrast of the interferometer by the factor $\int P(\delta v)\cos (2k\delta v t_\mathrm{sep}) \ d\delta v$, a convolution of the random phase shift with the probability distribution $P(\delta v)$. If we ignore the effects of vibrations and assume the contrast decay with wavepacket separation (set by $t_\mathrm{sep}$) is entirely caused by this mechanism, we can estimate an upper bound for this random velocity distribution. Using the measured $\kappa$ at $\tau_\mathrm{hold}=200\,T_\mathrm{BO,g}$ (inset of Fig. 2 in main text), we obtain a Lorentzian width of 40 $\mathrm{\mu m / s} = 0.01 v_\mathrm{rec}$ for $P(\delta v)$, far below our detection threshold. 

\section*{SIV. Coherence Time vs Temperature}

\begin{figure}
    \centering
    \includegraphics[width=\linewidth]{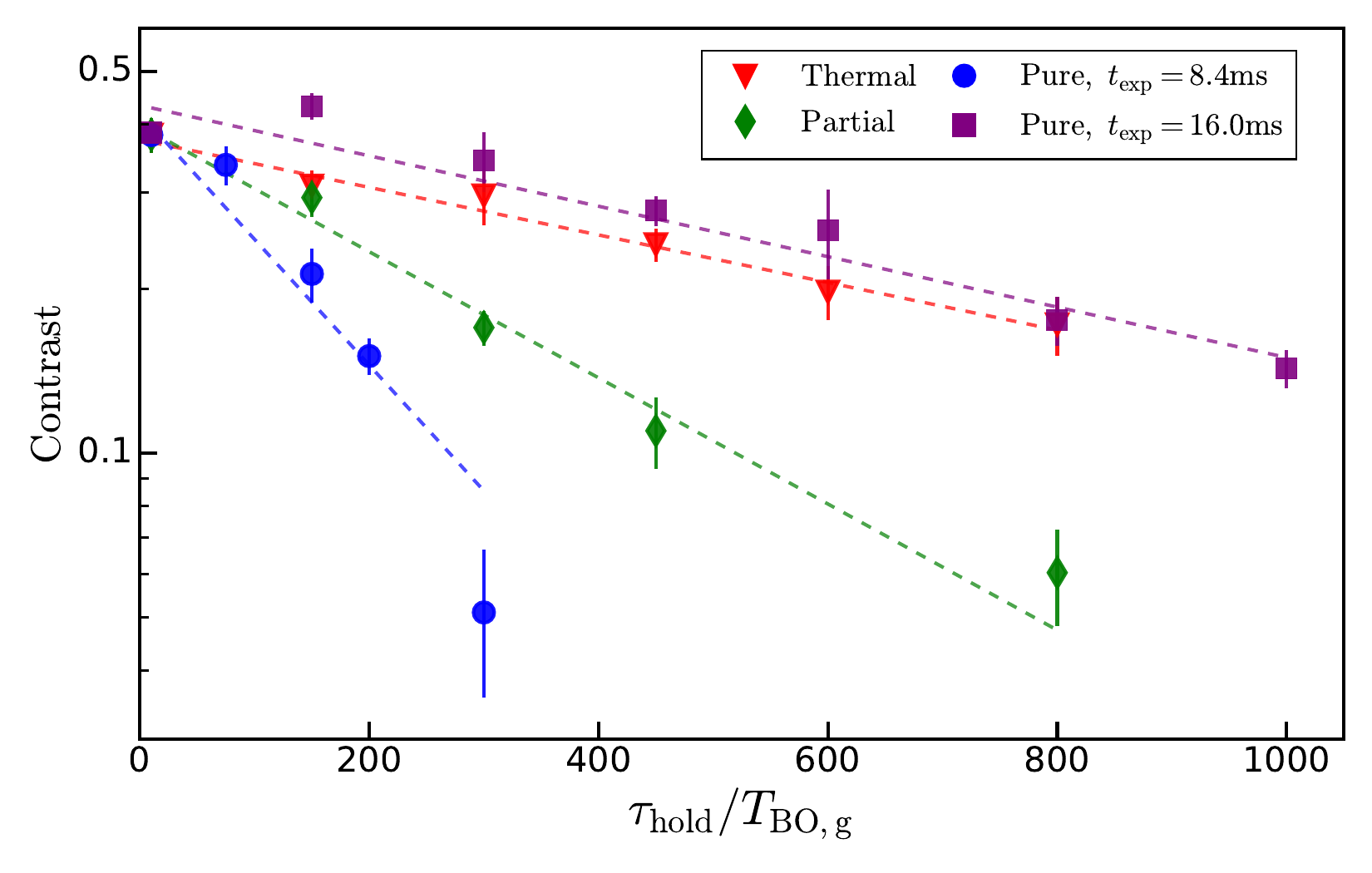}
    \caption{Coherence time vs temperature for a thermal cloud, partial condensate, and pure condensate with two expansion times, for a fixed wavepacket separation of $\Delta z / d = 15$. Dashed lines represent fits to an exponential decay with $1/e$ coherence times $\tau_C/T_\mathrm{BO,g}$ of 998(75), 376(36), 189(26), 941(113) for the thermal, partial, pure with $t_\mathrm{exp}=8.4\,\mathrm{ms}$, and pure with $t_\mathrm{exp}=16.0\,\mathrm{ms}$ respectively. Error bars represent $1\sigma$ fit uncertainties.}
    \label{fig:coherence_time_temp}
\end{figure}

Fig.~\ref{fig:coherence_time_temp} shows the coherence times for different temperature sources: thermal with $T=350\,$nK, partial condensate (where the evaporation sequence stops halfway between $T=350\,$nK and a pure BEC), and a pure BEC at two expansion times. At fixed expansion time, the increasing density from evaporative cooling increases the mean-field dephasing of the ensemble and thus decreases the coherence time (reported in caption). The coherence time of the expanded BEC is the same within statistical error of the thermal source, as discussed in Sec. V~A of the main text.

\section*{SV. Hold Time Fringe for Gravimetry}

As presented in Sec. V~C of the main text, the value of $g$ can be measured either through changing the laser phase $\Delta \phi_L$ and looking at the evolution of the fitted offset phase $\phi_0$, as presented in the main text, or through the frequency of the resulting fringe when changing $\tau_\mathrm{hold}$. For our Bragg-based interferometer, the change $\delta\tau_\mathrm{hold}$ around an integer number of BOs must be small with respect to the BO time $T_\mathrm{BO,g}$ to avoid systematic errors in the efficiency of $\pi/2$ pulses 3 and 4 due to Doppler shifts. The frequencies of the two beams appropriately compensate for the Doppler shift from the launch, but otherwise assume the Bragg transition is kept on resonance by the gravity chirp rate $\dot{\delta}_g$. If $\delta\tau_\mathrm{hold}$ becomes comparable to $T_\mathrm{BO,g}$, then the atoms will exit the lattice hold with a velocity on the order of the recoil velocity, introducing a significant Doppler shift which will affect both the contrast through the Bragg efficiency as well as the phase through the detuning.

\begin{figure}[b!]
    \centering
    \includegraphics[width=\linewidth]{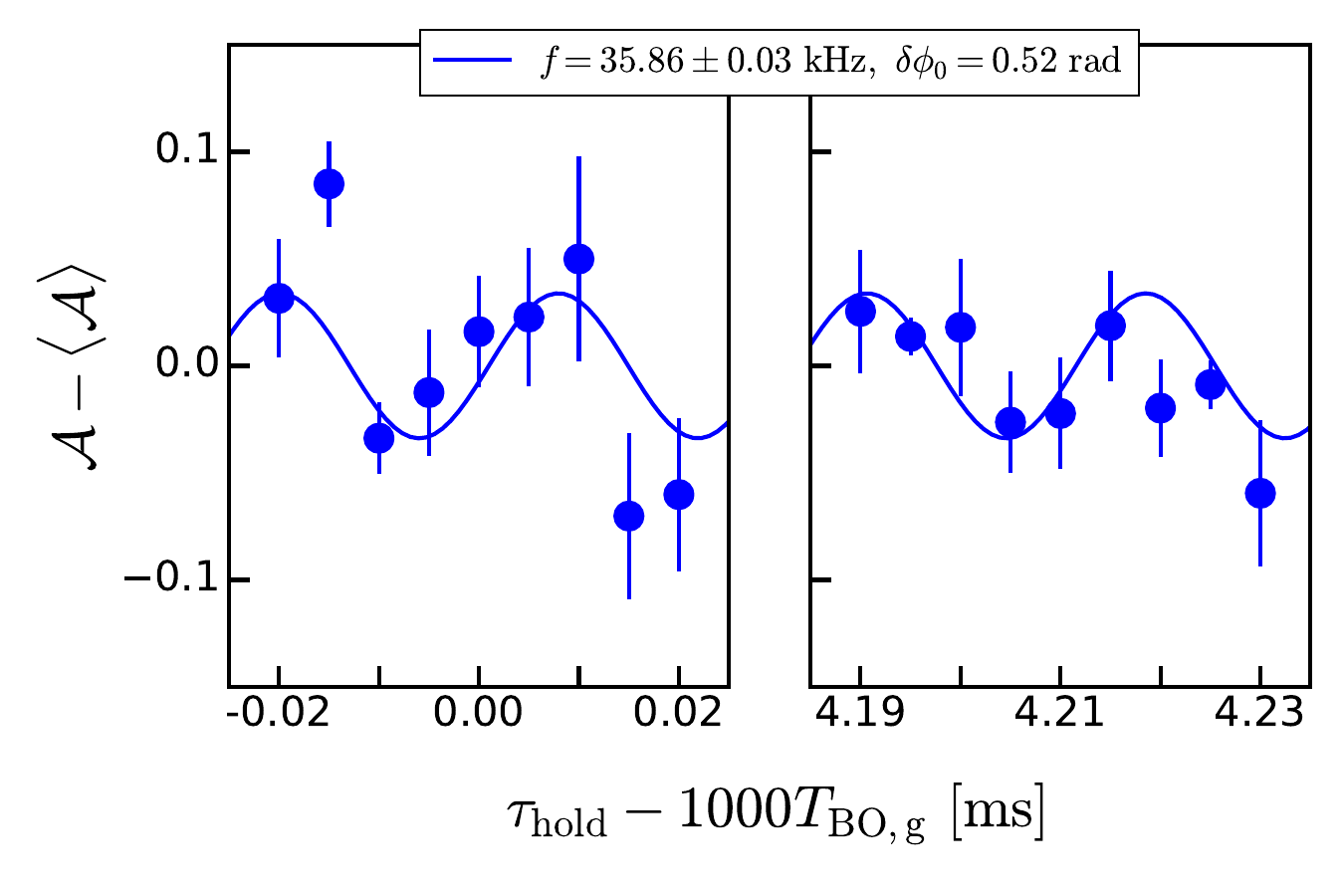}
    \caption{Demonstration of our trapped atom interferometer as a gravimeter through varying the hold time $\phi_\mathrm{prop}$. Error bars are the standard error on the mean with at least five points for each hold time. The value of $g$ affects the frequency of the fringe, which is consistent with a value of $9.81 \ \mathrm{m/s^2}$ and a separation of $\Delta z = 30 d \simeq 8.3 \ \mathrm{\mu m}$. In a bootstrapping scheme, the statistical uncertainty on the frequency (and thus $g$) is governed by the fractional frequency and phase uncertainty.}
    \label{fig:g_thold}
\end{figure}

For the example in Fig.~\ref{fig:g_thold}, taken at $\tau_\mathrm{hold}/T_\mathrm{BO,g}=1000$ with a wavepacket separation of $\Delta z = 30 d$, scanning approximately $3\pi$ of phase is accomplished with $|\delta\tau_\mathrm{hold}|\leq 20 \,\mathrm{\mu s}$. This is approximately a 2\% shift in the exit velocity, which is well within the spectral acceptance of our Bragg pulses.

Here, two fringes spaced by 5 BOs are fit to a global sinusoid for a fractional frequency uncertainty of 0.08\% with a value consistent with the nominal value of $g$. In a bootstrapping scheme to unwrap the phase, both the frequency and phase would be used with increasing hold times to build statistical precision. The fringe shown in Fig.~\ref{fig:g_thold} has a relative statistical uncertainty in the phase of approximately $3\times10^{-6}$.

\end{document}